\documentclass[aps,twocolumn,showpacs,amssymb,eqsecnum]{revtex4}
\usepackage{graphicx}
\usepackage[british]{babel}
\usepackage{amssymb}
\usepackage{amsmath}
\usepackage{amsfonts}
\usepackage[usenames]{color}
\usepackage{mathrsfs}
\begin{document}
\date{\today}
\title{Dynamics of DNA-breathing: Weak noise analysis, finite time
singularity, and mapping onto the quantum Coulomb problem}

\author{Hans C. Fogedby}
\email{fogedby@phys.au.dk} \affiliation{Department of Physics and
Astronomy, University of Aarhus\\Ny Munkegade, 8000, Aarhus C,
Denmark} \affiliation{Niels Bohr Institute for Astronomy, Physics,
and Geophysics\\ Blegdamsvej 17, 2100, Copenhagen {\O}, Denmark}
\author{Ralf Metzler}
\email{metz@ph.tum.de} \affiliation{ Physik Department, Technical
University of Munich, 85748 Garching, Germany}

\begin{abstract}
We study the dynamics of denaturation bubbles in double-stranded
DNA on the basis of the Poland-Scheraga model. We show that long
time distributions for the survival of DNA bubbles and the size
autocorrelation function can be derived from an asymptotic weak noise
approach. In particular, below the melting temperature the bubble
closure corresponds to a noisy finite time singularity.
We demonstrate that the associated
Fokker-Planck equation is equivalent to a quantum Coulomb problem. Below
the melting temperature the bubble lifetime is associated with the
continuum of scattering states of the repulsive Coulomb potential;
at the melting temperature the Coulomb potential vanishes and the
underlying first exit dynamics exhibits a long time power law
tail; above the melting temperature, corresponding to an
attractive Coulomb potential, the long time dynamics is controlled
by the lowest bound state. Correlations and finite size effects
are discussed.
\end{abstract}

\pacs{05.40.-a,02.50.-r,87.15.-v,87.10.+e}

\maketitle

\section{\label{intro}Introduction}

Under physiological conditions the Watson-Crick double-helix of DNA
constitutes the equilibrium structure, its stability ensured by
hydrogen-bonding of paired bases and base stacking between nearest
neighbor pairs of base pairs \cite{Kornberg74,Watson53}. By
variation of temperature or pH-value double-stranded DNA
progressively denatures, yielding regions of single-stranded DNA,
until the double-strand is fully molten. This is the helix-coil
transition taking place at a melting temperature $T_m$ defined as
the temperature at which half of the DNA molecule has undergone
denaturation \cite{Poland70}.

However, already at room temperature thermal fluctuations cause rare
opening events of small denaturation zones in the double-helix
\cite{Gueron87}. These \emph{DNA bubbles\/} consist of flexible
single-stranded DNA, and their size fluctuates in size by step-wise
zipping and unzipping of the base pairs at the two zipper forks
where the bubble connects to the intact double-strand. Below the
melting temperature $T_m$, once formed, a bubble is an intermittent
feature and will eventually zip close again. The multistate
\emph{DNA breathing\/} can be monitored in real time on the single
DNA level \cite{Altan-Bonnet03}. Biologically, the existence of
intermittent (though infrequent) bubble domains is important, as the
opening of the Watson-Crick base pairs by breaking of the hydrogen
bonds between complementary bases disrupts the helical stack. The
flipping out of the ordered stack of the unpaired bases allows the
binding of specific chemicals or proteins, that otherwise would not
be able to access the reactive sites of the bases
\cite{Gueron87,Poland70,Krueger06,Frank87}.

The size of the bubble domains varies from a few broken base pairs well
below $T_m$, up to some two hundred closer to $T_m$. Above $T_m$, individual
bubbles continuously increase in size, and merge with vicinal bubbles, until
complete denaturation \cite{Poland70}. Assuming that the
bubble breathing dynamics takes place on a slower time scale than
the equilibration of the DNA single-strand constituting the bubbles,
DNA-breathing can be interpreted as a random walk in the 1D
coordinate $x$, the number of denatured base pairs.

DNA breathing has been investigated in the Dauxois-Peyrard-Bishop
model \cite{Peyrard89,Dauxois93}, that describes the motion of
coupled oscillators representing the base pairs. On the basis of the
Poland-Scheraga model,
DNA breathing has been studied in terms of continuous
Fokker-Planck approaches \cite{Hwa03,Hanke03}, and in terms of the
discrete master equation and the stochastic Gillespie scheme
\cite{Banik05,Ambjorn05,Ambjorn06,Ambjorn07,Ambjorn07a,Bicout04}.
The coalescence of two bubble domains was analyzed in
Ref.~\cite{Novotny07}.

In what follows we study the Langevin and Fokker-Planck
non-equilibrium extension of the Poland-Scheraga model in terms of
both a general weak noise approach accessing the long time behavior,
see e.g., Refs.~\cite{Fogedby99a,Fogedby03b}, and a mapping to a
quantum Coulomb problem \cite{Fogedby07}. This allows us to
investigate in more detail the finite time singularity underlying
the breathing dynamics, as well as the survival of individual
bubbles. The paper is organized in the following manner. In
Sec.~\ref{model}, we introduce and discuss the model, in
Sec.~\ref{weak} we apply the weak noise approach and extract long
time results and study the stability of the solutions. In
Sec.~\ref{coulomb} we map the problem to a quantum Coulomb problem
and derive the long-time scaling of the bubble survival. Finally, in
Sec.~\ref{discussion} we discuss the results and draw our
conclusions in Sec.~\ref{summary}.
\section{\label{model}Dynamic model for DNA breathing}
In the Poland-Scheraga free energy approach, bubbles are introduced as
free energy changes to the double-helical ground state, such that the
disruption of each additional base pair of a bubble requires to cross
an energetic barrier that is rewarded by an entropy gain. While the
persistence length of double-stranded DNA is rather large (of the order
of 50nm) and it is assumed to have no configurational entropy, the
single-stranded bubbles are flexible, and therefore behave like a
polymer ring. The Poland-Scheraga partition factor
for a single bubble in a homopolymer is of the form
\begin{equation}
\mathcal{Z}(m)=\sigma_0u^m(1+m)^{-c},
\end{equation}
where $m$ counts the (discrete) number of broken base pairs, and
$u=\exp\left( -\beta\gamma\right)$, with $\beta=1/[kT]$, is the
Boltzmann factor for breaking the stacking interactions when
disrupting an additional base pair. The cooperativity factor
$\sigma_0=\exp\left(-\beta\gamma_0\right)$ quantifies the
so-called boundary energy $\gamma_0$ for initiating a bubble.
$\gamma_0$ is of the order of 8000 cal/mol, corresponding to
approximately 13 $kT$ at $37^\circ$C. Occasionally, somewhat
smaller values for $\sigma_0$ are assumed, down to approximately 8 $kT$.
Bubbles below the melting point of DNA are therefore rare events.
Typical equilibrium melting temperatures of DNA for standard salt conditions
are in the range $T_m\sim 70-100^{\circ}$C, depending on the
relative content of weaker AT and stronger GC Watson-Crick base
pairs. Thus, double-stranded DNA denatures at much higher
temperatures as many proteins. Note that the melting temperature
of DNA can also be increased by change of the natural winding, as
opening of the double-strand in ring DNA is coupled with the
creation of superstructure; this is the case, for instance, in
underwater bacteria living in hot vents, compare Ref.~\cite{ctn},
and references therein.

Due to the large value of $\sigma_0$, below the melting temperature
to good approximation individual bubbles are statistically
independent, and therefore a one-bubble picture appropriate. Having
experimental setups in mind as realized in
Ref.~\cite{Altan-Bonnet03}, where special DNA constructs are
designed such that they have only one potential bubble domain, we
also consider a one-bubble picture at and above $T_m$. Our results
are meant to apply to such typical single molecule setups. In
comparison to the rather high energy barrier $\gamma_0$, according
to which the opening of a bubble corresponds to a nucleation
process, to break the stacking of a single pair of base pairs
requires much less thermal activation, ranging from $\gamma=-0.1$ to $+3.9$
$kT$ for TA/AT and GC/CG pairs of base pairs at $37^\circ$C,
respectively; here, the positive sign refers to a thermodynamically
stable state. These comparatively low values for the stacking free energy of
base pairs stems from the fact that stacking enthalpy cost and entropy
release on base pair disruption almost cancel.
Finally, the term $(1+m)^{-c}$ measures the entropy
loss on formation of a closed polymer ring, with respect to a linear
chain of equal length. The offset by 1 is often taken into account
to represent the short persistence length of single stranded DNA.
For the critical exponent $c$, one typically uses the value $1.76$
of a Flory chain in three dimensions
\cite{Wartell85,Poland66,santalucia,blake,Krueger06,Ambjorn06}, while
a slightly larger value ($c=2.12$) was suggested based on different
polymer models
\cite{Guttmann04,Carlon02,Bar07,Kafri00,Kafri02,monthus}.
Here, we disregard the offset, and consider the pure
power-law form $m^{-c}$.

In the following, we consider the continuum limit of the above picture,
measuring the "number" of broken base pairs with the continuous variable
$x$. The Poland-Scheraga free energy for a single bubble then has the
form \cite{Poland70,Hanke03}
\begin{eqnarray}
\label{fe}
\mathscr{F}=\gamma_0+\gamma x+ckT\ln x. \label{free}
\end{eqnarray}
where $x\ge 0$ is the bubble size as measured in units of base pairs.
Treating the bubble size $x$ as a continuum variable, we impose an
absorbing wall at $x=0$, the zero-size bubble. The completely closed
bubble state is stabilized by the size of the cooperativity factor
$\sigma_0$, and bubbles therefore become rare events. Expression (\ref{fe})
corresponds to a logarithmic sink in $\mathscr{F}$ at $x=0$. The
free energy density $\gamma(T)$ has a temperature dependence, which we
write as
\begin{eqnarray}
\gamma(T)=\gamma_1(T_m-T)/T_m,\label{gam}
\end{eqnarray}
where $T_m$ is the melting temperature.

From Eq. (\ref{free}) it follows that a characteristic bubble size
is set by $x_1=ckT/|\gamma|$. For large bubble size $x>x_1$ the
linear term dominates and the free energy grows like $\mathscr{F}
\sim\gamma_0+\gamma x$. For small bubbles $x<x_1$ [or close to
$T_m$, where $\gamma(T)\approx0$] the free energy is characterized
by the logarithmic sink but has strictly speaking a minimum at
$\mathscr{F}=\gamma_0$ for zero bubble size. We distinguish two
temperature ranges:

(i) For $\gamma<0$, i.e.,
$T>T_m$, the free energy has a maximum
$\mathscr{F}_{\text{max}}=\gamma_0+ckT(\log x_1-1)$ at $x=x_1$.
The free energy profile thus defines a Kramers escape problem in
the sense that an initial bubble can grow in size corresponding to
the complete denaturation of the double stranded DNA. The escape
probability $P_{\text{esc}}\propto\exp(-\Delta\mathscr{F}/kT)$,
where the free energy barrier is $\Delta\mathscr{F}=ckT(\log x_1 -
1)$, i.e.,
\begin{eqnarray}
P_{\text{esc}}\propto\left(\frac{ckT}{|\gamma|}\right)^{-c}.
\label{escape}
\end{eqnarray}

(ii) For $\gamma>0$, i.e., $T<T_m$, the free energy increases
monotonically from $\mathscr{F}=\gamma_0$ at $x=0$ and the finite
size bubbles are stable. The change of sign of $\gamma$ at $T=T_m$
thus defines the bubble melting.

\begin{figure}
\includegraphics[width=0.8\hsize]{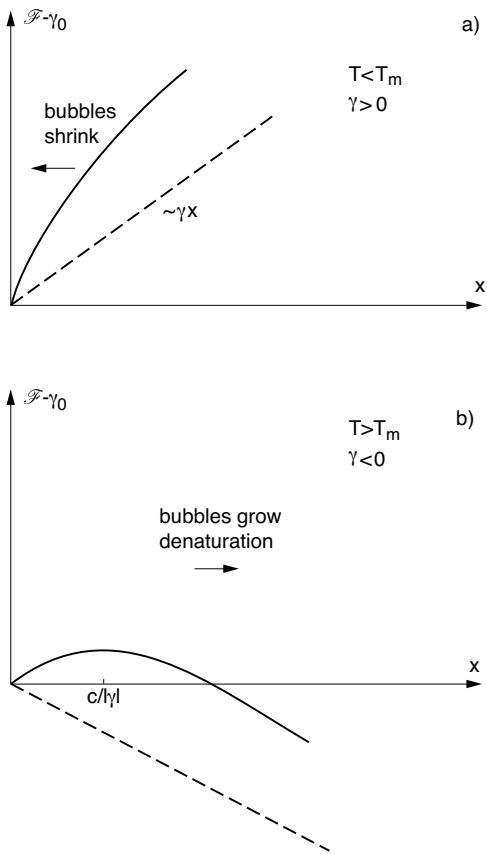}
\caption {We depict the free energy profile $\mathscr{F}-\gamma_0$
below and above the melting temperature $T_m$ as a function of
bubble size. In a) we show $\mathscr{F}-\gamma_0$ for $\gamma>0$,
i.e., $T<T_m$; in b) we show $\mathscr{F}-\gamma_0$ for
$\gamma<0$, i.e., $T>T_m$. For large bubble sizes, $x\gg x_1$ the
free energy behaves approximately linearly as function of bubble
size. For small bubble sizes the free energy has a logarithmic
sink corresponding to the absorbing state at $x=0$ (arbitrary
units). Above melting, there exists a nucleation barrier that
needs to be crossed before the bubble is allowed to grow towards
full denaturation. In both cases, the comparatively high initiation
barrier $\gamma_0$ has to be overcome to seed the bubble.} \label{fig1}
\end{figure}

For $\gamma<0$, i.e., $T>T_m$,
the free energy has a maximum and decreases for large bubble size,
as a result the bubbles expand and the double stranded DNA
denatures, that is, melts. In Fig.~\ref{fig1} we have depicted
the free energy profile as a function of bubble size for
$\gamma>0$, $T<T_m$, and for $\gamma<0$, $T>T_m$.

The stochastic bubble dynamics in the free energy landscape $\mathscr{F}$ is
described by the Langevin equation
\begin{eqnarray}
\frac{dx}{dt}=-D\frac{d\mathscr{F}}{dx}+\xi, \label{lan1}
\end{eqnarray}
driven by thermal noise $\xi$, that is characterized by the correlation
function
\begin{eqnarray}
\langle\xi(t)\xi(t')\rangle=2DkT\delta(t-t'). \label{noise}
\end{eqnarray}
The kinetic coefficient $D$ of dimension
$(kT)^{-1}s^{-1}$ sets the inverse time scale of the
dynamics. Inserting the free energy (\ref{free}) in Eq.
(\ref{lan1}) we have in particular
\begin{eqnarray}
\frac{dx}{dt}=\Omega_2 -\frac{\Omega_1}{x}+\xi, \label{lan2}
\end{eqnarray}
where we have found it convenient to introduce the inverse time
scales $\Omega_1$ and $\Omega_2$,
\begin{subequations}
\begin{eqnarray}
&&\Omega_1=DckT, \label{inv1}
\\
&&\Omega_2=-D\gamma=D\gamma_1(T-T_m)/T_m.\label{inv2}
\end{eqnarray}
\end{subequations}
Note that the characteristic bubble size $x_1=ckT/|\gamma|$ is given
by
\begin{eqnarray}
x_1=\frac{ckT}{\gamma}=\frac{\Omega_1}{|\Omega_2|} \label{size}
\end{eqnarray}
and thus emerges from the time scale competition between the
$\Omega_i$, from a dynamic point of view.

In the limits of large and small bubble sizes, the Langevin equation
(\ref{lan1}) allows exact solutions:

(i) For large bubble size $x\gg x_1$ we can ignore the loop closure or
entropic contribution $ckT/x$ and we obtain the Langevin equation
\begin{eqnarray}
\frac{dx}{dt}=\Omega_2+\xi, \label{lan3}
\end{eqnarray}
describing a 1D random walk with an overall drift velocity
$\Omega_2$. For large $x$ we thus obtain the distribution
\cite{Risken89}
\begin{eqnarray}
P(x,t)=\frac{1}{\sqrt{4\pi DkTt}}\exp\left[-\frac{(x-x_0-\Omega_2
t)^2}{4DkTt}\right],~~ \label{dis1d}
\end{eqnarray}
where $x_0$ is the initial (large) bubble size. It follows that
the mean bubble size scales linearly with time, $\langle x\rangle
= x_0+\Omega_2t$. Below $T_m$ ($\Omega_2<0$) the bubble size
shrinks towards bubble closure; above $T_m$ ($\Omega_2>0$) the
bubble size grows, leading to denaturation. The mean square bubble
size fluctuations $\langle(\Delta x)^2\rangle=2DkTt$, increase
linearly in time, a typical characteristic of a random walk.

Taking into account the absorbing state condition $P(x=0,t)=0$ for
zero bubble size by forming the linear combination (method of
images), we obtain for the distribution \cite{Redner01}
\begin{eqnarray}
\nonumber
&&\hspace*{-0.4cm}
P_{\text{abs}}=\frac{1}{\sqrt{4\pi DkTt}}\left(\exp\left\{-\frac{(x-x_0-
\Omega_2t)^2}{4DkTt}\right\}\right.\\
&&\left.-\exp\left\{-\frac{x_0\Omega_2}{DkT}\right\}\exp\left\{-\frac{(x+x_0
-\Omega_2t)^2}{4DkTt}\right\}\right),\hspace*{0.4cm}
\label{absdis}
\end{eqnarray}
and infer, using the definition \cite{Redner01}
\begin{eqnarray}
W(t)=-\int_0^\infty dx \frac{\partial P_{\text{abs}}}{\partial t},
\label{def}
\end{eqnarray}
the first passage time density
\begin{eqnarray}
W(t)=\frac{x_0}{\sqrt{4\pi DkTt^3}}\exp\left(-\frac{(x_0+\Omega_2t)^2}{4DkTt}
\right).
\label{abs}
\end{eqnarray}
with the typical Sparre Andersen asymptotics
\begin{equation}
W(t)\sim\frac{x_0}{\sqrt{4\pi DkT}}t^{-3/2}.
\end{equation}

\begin{figure}
\includegraphics[width=0.9\hsize]{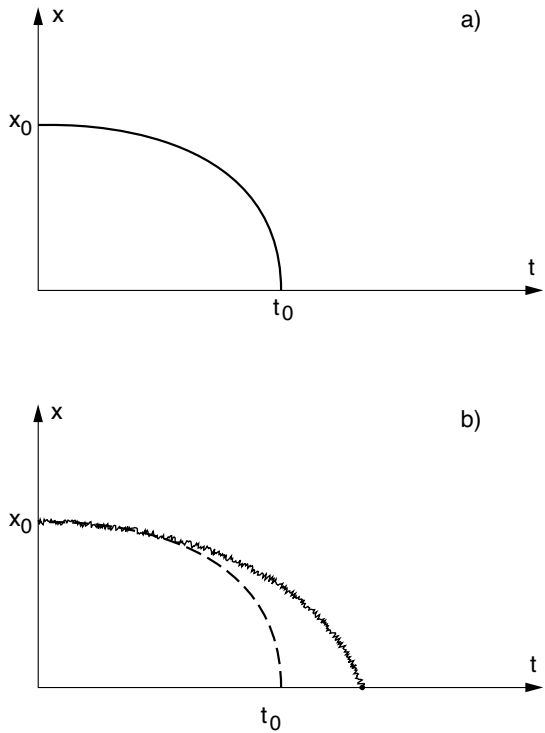}
\caption{In a) we show the time
evolution of a small bubble of size $x$ in the absence of thermal
noise. For $x=0$ corresponding to bubble closure we encounter a
finite-time-singularity at $t_0=x_0/2\Omega_1$. In b) we depict
the noisy case. Here the first passage time is a statistical event
characterized by $W(t)$ (arbitrary units).} \label{fig2}
\end{figure}

(ii) For small bubble size $x\ll x_1$ the nonlinear entropic term
dominates and the bubble dynamics is governed by the nonlinear
Langevin equation
\begin{eqnarray}
\frac{dx}{dt}=-\frac{\Omega_1}{x}+\xi. \label{lan4}
\end{eqnarray}
For vanishing noise Eq. (\ref{lan4}) has the solution
$x=(2\Omega_1)^{1/2}(t_0-t)^{1/2}$ with $t_0=x_0/2\Omega_1$ in
terms of the initial bubble size $x_0$ and thus exhibits a finite
time singularity for $x=0$, i.e., a zero bubble size or bubble
closure at time $t_0$. In Fig.~\ref{fig2} we have depicted the
finite-time-singularity solution for vanishing noise together with
the noisy case.

In the presence of thermal noise Eq. (\ref{lan4}) admits an exact
solution, see e.g. Ref. \cite{Fogedby02d}. The probability
distribution, subject to the absorbing state condition $P(0,t)=0$,
has the form
\begin{eqnarray}
P(x,t)&=&\frac{x^{\Omega_1/2DkT+1/2}}{x_0^{\Omega_1/DkT-1/2}}
\frac{e^{-(x^2+x_0^2)/4DkTt}}{2DkTt}
\nonumber
\\
&&\times I_{1/2+\Omega_1/2DkT}\left(\frac{xx_0}{2DkTt}\right).
\label{absdis2}
\end{eqnarray}
Here $I_\nu$ is the Bessel function of imaginary argument,
$I_\nu(z)=(-i)^\nu J_\nu(iz)$ \cite{Lebedev72}. Correspondingly,
we find the first passage time distribution
\begin{eqnarray}
\nonumber
W(t)=&&\frac{4DkTx_0^{1+\Omega_1/DkT}}{\Gamma(1/2-\Omega_1/2DkT)}
\exp\left(-\frac{x_0^2}{4DkTt}\right)\\
&&\times (4DkTt)^{-3/2-\Omega_1/2DkT}
\label{abs2}
\end{eqnarray}
with the long time tail
\begin{equation}
W(t)\sim\frac{x_0^{1+\Omega_1/DkT}t^{-3/2-c/2}}{
\Gamma(1/2-\Omega_1/2DkT)(4DkT)^{1/2+\Omega_1/2DkT}},
\end{equation}
where we substituted back for $\Omega_1$: For small bubble sizes, the
exponent $c$ due to the polymeric interactions changes the first passage
statistics. As already noted in Ref.~\cite{Fogedby07}, this modified
exponent for $c>1$ gives rise to a finite mean first passage time
$\int_0^{\infty}tW(t)dt$, in contrast to the first passage time distribution
(\ref{abs})

In the general case for bubbles of all sizes the fluctuations of
double-stranded DNA is described by Eq.~(\ref{lan2}). The associated
Fokker-Planck equation for the distribution $P(x,t)$ has the form (compare
also Refs.~\cite{Hanke03,Bar07,Fogedby07})
\begin{eqnarray}
\frac{\partial P}{\partial t}=\frac{\partial}{\partial
x}\left(-\Omega_2+\frac{\Omega_1}{x}\right)P+
DkT\frac{\partial^2P}{\partial x^2},\label{fp}
\end{eqnarray}
and provides the complete description of the single bubble
dynamics in double-stranded homopolymer DNA in the continuum limit
of the Poland-Scheraga model. For large bubble sizes where the
entropic term $\Omega_1/x$ can be neglected the solution of
Eq.~(\ref{fp}) is given by Eqs.~(\ref{dis1d}) and (\ref{absdis}).
Conversely, for small bubble sizes, where the entropic term
$\Omega_1/x$ dominates, or for all bubble sizes precisely at the
transition temperature $\Omega_2=0$ ($T=T_m$), the solution of
Eq.~(\ref{fp}) is given by the noisy finite-time-singularity
solution in Eqs.~(\ref{absdis2}) and (\ref{abs2}).
\section{\label{weak}Weak noise analysis}
In the weak noise limit $DkT\rightarrow 0$ we can apply a
well-established canonical scheme to investigate the Fokker-Planck equation
(\ref{fp}), see, for instance, Refs.~\cite{Fogedby99a,Fogedby03b}.
Introducing the WKB ansatz
\begin{eqnarray}
P(x,t)\propto\exp\left(-\frac{S(x,t)}{2DkT}\right),\label{wkb}
\end{eqnarray}
the weight (or action) $S(x,t)$ satisfies the Hamilton-Jacobi equation
\begin{equation}
\frac{\partial S}{\partial t}+H=0
\end{equation}
with Hamiltonian
\begin{eqnarray}
H=\frac{1}{2}p^2-p\left(-\Omega_2+\frac{\Omega_1}{x}\right).\label{ham1}
\end{eqnarray}
From this scheme, the equations of motion yield in the form
\begin{eqnarray}
&&\frac{dx}{dt}=\left(\Omega_2-\frac{\Omega_1}{x}\right)+p,\label{eq1}
\\
&&\frac{dp}{dt}=-\frac{\Omega_1}{x^2}p.\label{eq2}
\end{eqnarray}
They determine orbits in a canonical phase space spanned by the bubble
size $x$ and the momentum $p$. Comparing the equation of motion
(\ref{eq1}) with the Langevin equation (\ref{lan2}) we observe that
the thermal noise $\xi$ is replaced by the momentum $p=\partial
S/\partial x$.

The action $S$ associated with an orbit
from $x_0$ to $x$ during time $t$ is given by
\begin{eqnarray}
S(x,t)=\int_{x_0,0}^{x,t}dt~p\frac{dx}{dt}-Ht, \label{act1}
\end{eqnarray}
or by insertion of Eq. (\ref{eq1})
\begin{eqnarray}
S(x,t)=\frac{1}{2}\int_{x_0,0}^{x,t}dt~p^2. \label{act2}
\end{eqnarray}
%
\subsection{Large bubbles}
\begin{figure}
\includegraphics[width=1.0\hsize]{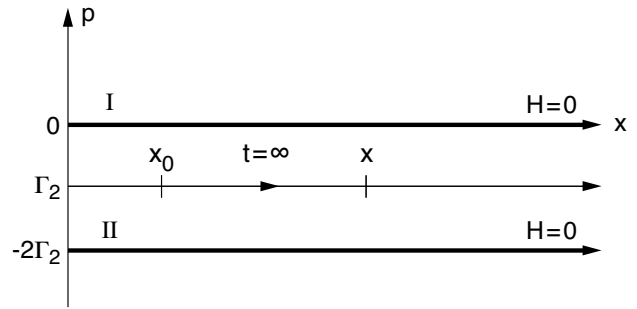}
\caption {We show the phase space
structure in the case $\Omega_1=0$, i.e., for random walk with
constant drift. We show the zero energy manifolds for $p=0$ and
$p=-2\Omega_2$ and a negative energy orbit from $x_0$ to $x$ in
time $t$ (arbitrary units).} \label{fig3}
\end{figure}

For large bubbles, i.e., $x\gg x_1=\Omega_1/|\Omega_2|$, we can
ignore the loop closure contribution characterized by  $\Omega_1$,
and we obtain the Hamiltonian
\begin{eqnarray}
H=\frac{1}{2}p^2+\Omega_2p, \label{ham2}
\end{eqnarray}
as well as the linear equations of motion
\begin{eqnarray}
&&\frac{dx}{dt}=\Omega_2+p,\label{eq10}
\\
&&\frac{dp}{dt}= 0.\label{eq20}
\end{eqnarray}
The solution is given by $p=p_0$, $x=x_0+(p_0+\Omega_2)t$ describing an orbit
from $(x_0,p_0)$ to $(x,p_0)$ in time $t$. Isolating
$p_0=(x-x_0-\Omega_2)/t$ and inserting in Eq. (\ref{act2}) we obtain
the action
\begin{eqnarray}
S(x,t)=\frac{1}{2}\frac{(x-x_0-\Omega_2 t)^2}{t}, \label{act3}
\end{eqnarray}
and inserted in Eq. (\ref{wkb}) the biased random walk distribution
(\ref{dis1d}).  In Fig.~\ref{fig3} we have depicted the phase space
for $\Omega_1=0$, i.e., in the large bubble-random walk case. The
orbits are confined to the constant energy surfaces. We note in
particular that the infinite time orbit lies on the $p=-\Omega_2$
manifold. We note, moreover, that in the large bubble case the weak
noise case fortuitously yields the exact result for the distribution
$P$.
%

\subsection{Small bubbles at and below $T_m$}
For small bubbles, i.e.,  $x\ll x_1=\Omega_1/|\Omega_2|$, the loop
closure contribution dominates and we obtain the Hamiltonian
\begin{eqnarray}
H=\frac{1}{2}p^2-\frac{p\Omega_1}{x}, \label{ham3}
\end{eqnarray}
and the equations of motion
\begin{eqnarray}
&&\frac{dx}{dt}=-\frac{\Omega_1}{x}+p,\label{eq21}
\\
&&\frac{dp}{dt}=-\frac{\Omega_1}{x^2}p,\label{eq22}
\end{eqnarray}
determining orbits in $(x,p)$ phase space. Eliminating $p$ the
bubble size is governed by the second order equation
\begin{eqnarray}
&&\frac{d^2x}{dt^2}=-\frac{dV}{dx},\label{eqA}
\\
&&V=-\frac{\Omega_1^2}{2x^2},\label{eqB}
\end{eqnarray}
describing the 'fall to the center' ($x=0$) of a bubble of size $x$, i.e.,
the absorbing state corresponding to bubble closure.

\begin{figure}
\includegraphics[width=1.0\hsize]{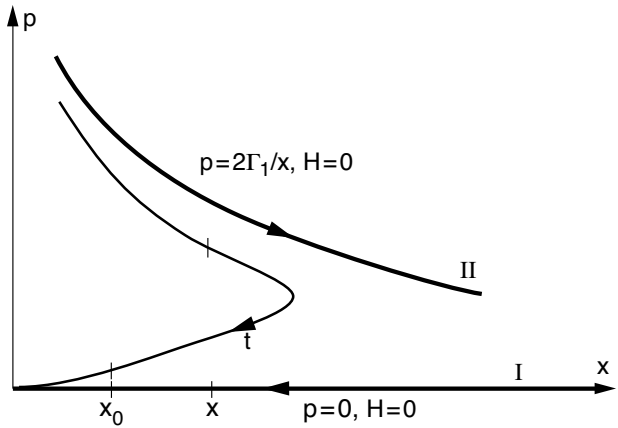}
\caption{We show the phase space structure in the case $\Omega_2=0$ ($T=T_m$),
i.e., for the small bubble
dynamics governed by the entropic contribution. We show the zero
energy manifolds $p=0$ and $p=2\Omega_1/x$ and a negative energy
orbit from $x_0$ to $x$ in time $t$ (arbitrary units).}
\label{fig4}
\end{figure}

The long time stochastic dynamics is here governed by the
structure of the zero energy manifolds and fixed points. From Eq.
(\ref{ham3}) it follows that the zero energy manifold has two
branches: i) $p=0$, corresponding to the noiseless transient
behavior showing a finite time singularity as depicted in
Fig.~\ref{fig2} and ii) $p=2\Omega_1/x$ associated with the noisy
behavior. In Fig.~\ref{fig4} we have depicted the phase space
structure.
In the long time limit the orbit from $x_0$ to $x$ passes close to
the zero energy manifold $p=2\Omega_1/x$. Inserted in the equation
of motion (\ref{eq21}) we have
\begin{eqnarray}
\frac{dx}{dt}=\frac{\Omega_1}{x}, \label{eqm1}
\end{eqnarray}
with long time solution
\begin{eqnarray}
x(t)\sim (2\Omega_1t)^{1/2}. \label{sol1}
\end{eqnarray}
We notice that the motion on the noisy manifold $p=2\Omega_1/x$ is
time reversed of the motion on the noiseless manifold $p=0$. Next
inserting the zero energy manifold condition $p=2\Omega_1/x$ in Eq.
(\ref{act2}) we obtain
\begin{eqnarray}
S = 2\Omega_1^2\int dt\left(\frac{1}{x}\right)^2, \label{act44}
\end{eqnarray}
and inserting the solution in Eq. (\ref{sol1}) the action
\begin{eqnarray}
S(x,t)= 2\Omega_1 \log x(t), \label{act4}
\end{eqnarray}
yielding according to Eq. (\ref{wkb}) the long time distribution
\begin{eqnarray}
P(x,t)\propto x(\Omega_1 t)^{-\Omega_1/2DkT}.\label{abdis3}
\end{eqnarray}
We have incorporated the absorbing state condition $P=0$ for $x=0$;
as discussed in Ref. \cite{Fogedby02d} this condition follows from
carrying the WKB weak noise approximation to next asymptotic order.
For the first-passage time density of loop closure we obtain correspondingly
\begin{eqnarray}
W(t)\propto t^{-\Omega_1/2DkT}.\label{abs3}
\end{eqnarray}
We note that the power law dependence in Eqs. (\ref{abdis3}) and
(\ref{abs3}) is in accordance with Eqs. (\ref{absdis2}) and
(\ref{abs2}) for $DkT\rightarrow 0$.
\section{\label{coulomb}Case of arbitrary noise strength}
In the previous section we inferred weak noise-long time expressions
for the distribution $P$ on the basis of a canonical phase space
approach. Here we address the Fokker-Planck equation (\ref{fp}) in
the general case. For the purpose of our discussion it is useful to
introduce the parameters
\begin{subequations}
\begin{eqnarray}
&&\mu=c/2, \label{par1}
\\
&&\epsilon=\frac{\gamma_1}{2k}\left(\frac{1}{T_m}-\frac{1}{T}\right).
\label{par2}
\end{eqnarray}
\end{subequations}
Measuring time in units of $\mu\text{s}$ the Fokker-Planck
equation (\ref{fp}) takes on the reduced form
\begin{eqnarray}
\frac{\partial P}{\partial t}=\frac{\partial}{\partial
x}\left(\frac{\mu}{x}-\epsilon\right)P+
\frac{1}{2}\frac{\partial^2P}{\partial x^2}.\label{fp2}
\end{eqnarray}
Note that $\mu\approx 1$, and, close to the physiological temperature
$T_{\text{r}}$, $\epsilon\approx 2(T/T_m-1)$.
\subsection{Connection to the quantum Coulomb problem}
\begin{figure}
\includegraphics[width=1.\hsize]{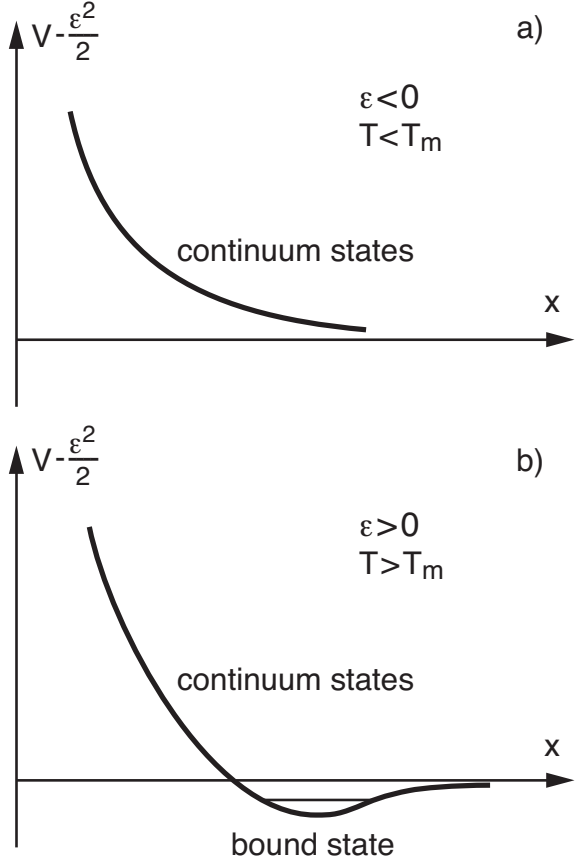}
\caption{Schematic of the potential $V(x)-\epsilon^2/2$. a)
$T<T_m$: The potential is repulsive, yielding a continuous
spectrum. The bubble fluctuations correspond to a biased Brownian
walk process in bubble size $x$ before collapse at $x=0$. b)
$T<T_m$. The potential is attractive and can trap a series of
bound states. At long times the lowest bound state indicated in
the figure controls the behavior. The bubbles increase in size
eventually leading to complete denaturation.} \label{fig5}
\end{figure}

By means of the substitution $P=e^{\epsilon x}x^{-\mu}\tilde P$,
$\tilde P$ satisfies the equation \cite{Fogedby07}
\begin{equation}
-\frac{\partial\tilde P}{\partial t}=
-\frac{1}{2}\frac{\partial^2\tilde P}{\partial x^2}+
\left(\frac{\mu(\mu+1)}{2x^2}-
\frac{\mu\epsilon}{x}+\frac{\epsilon^2}{2}\right)\tilde P,
\label{fp3}
\end{equation}
which can be identified as an imaginary time Schr\"{o}dinger
equation for a particle with unit mass in the potential
\begin{equation}
V(x)=\frac{\mu(\mu+1)}{2x^2}-\frac{\mu\epsilon}{x}+\frac{\epsilon^2}{2},
\label{pot}
\end{equation}
i.e., subject to the centrifugal barrier $\mu(\mu+1)/x^2$ for an
orbital state with angular momentum $\mu$ and a Coulomb potential
$-\mu\epsilon/x$. In Fig.~\ref{fig5} we have depicted the
potential $V-\epsilon^2/2$ in the two cases.
In terms of the Hamiltonian
\begin{equation}
H=-\frac{1}{2}\frac{d^2}{dx^2}+\frac{\mu(\mu+1)}{2x^2}-
\frac{\mu\epsilon}{x}+\frac{\epsilon^2}{2}, \label{ham4}
\end{equation}
the eigenvalue associated with Eq.~(\ref{fp3}) problem has the form
\begin{equation}
H\Psi_n=E_n\Psi_n. \label{eigen}
\end{equation}
Expressed in terms of the eigenfunctions the transition
probability $P(x,x_0,t)$ then becomes
\begin{eqnarray}
P(x,x_0,t)=e^{\epsilon(x-x_0)}\left(\frac{x_0}{x}\right)^{\mu}
\sum_ne^{-E_nt}\Psi_n(x)\Psi_n(x_0).\nonumber \\ \label{dis}
\end{eqnarray}
Here, the completeness of $\Psi_n$ ensures the initial condition
$P(x,x_0,0)=\delta(x-x_0)$. Moreover, in order to account for the
absorbing boundary condition for vanishing bubble size we choose
$\Psi_n(0)=0$. We also note that for a finite strand of length
$L$, i.e., a maximum bubble size of $L$, we have in addition the
absorbing condition $\Psi_n(L)=0$ for complete denaturation.
Expression (\ref{dis}) is the basis for our discussion of
DNA-breathing, relating the dynamics to the spectrum of
eigenstates, i.e., the bound and scattering states of the
corresponding Coulomb problem \cite{Landau59c}.

The transition probability $P(x,x_0,t)$ for the occurrence of a
DNA bubble of size $x$ at time $t$ is controlled by the Coulomb
spectrum. Below the melting temperature $T_m$ for
$\epsilon\propto(T/T_m-1)<0$, the Coulomb problem is repulsive and
the states form a continuum, corresponding to a random walk in
bubble size terminating in bubble closure $(x=0)$. At the melting
temperature $T_m$ for $\epsilon=0$, the Coulomb potential is
absent and the continuum of states is governed by the centrifugal
barrier alone, including the limiting case of a regular random
walk. Above the melting temperature for $\epsilon>0$, the Coulomb
potential is attractive and can trap an infinity of bound states;
at long times it follows from Eq. (\ref{dis}) that the lowest
bound state in the spectrum dominates the bubble dynamics,
corresponding to complete denaturation of the DNA chain.

Mathematically, we model the bubble dynamics with absorbing boundary
conditions at zero bubble size $x=0$, and, for a finite chain of length
$L$, at $x=L$. When the bubble vanishes or complete denaturation is
reached, that is, the dynamics stops. Physically, this stems from the
observation that on complete annihilation (closure) of the bubble, the
large bubble initiation barrier prevents immediate reopening of the bubble.
Similarly, a completely denatured DNA needs to re-establish bonds between
bases, a comparatively slow diffusion-reaction process.
\subsubsection{Long times for $T<T_m$}
At long times and fixed $x$ and $x_0$, it follows from Eq.
(\ref{dis}) that the transition probability is controlled by the
bottom of the energy spectrum. Below and at $T_m$  the spectrum is
continuous with lower bound $\epsilon^2/2$. Setting
$E_k=\epsilon^2/2+k^2/2$ in terms of the wavenumber $k$ and noting
from the eigenvalue problem in Eqs. (\ref{ham4}) and (\ref{eigen})
that $\Psi_k(x)\sim (kx)^{1+\mu}$ for small $kx$ we find
\begin{eqnarray}
\nonumber
P(x,x_0,t)&\propto&
\exp\left(-|\epsilon|(x-x_0)\right)\left(\frac{x_0}{x}\right)^{\mu}
\exp\left(-\frac{\epsilon^2t}{2}\right)\\
&&\times
\int_0^\infty dk e^{-k^2t/2}(k^2xx_0)^{1+\mu}.
\nonumber\\
\label{dis2}
\end{eqnarray}
By a simple scaling argument we then obtain the long time expression
for the probability distribution
\begin{eqnarray}
P(x,x_0,t)\propto xx_0^{1+2\mu}
e^{-|\epsilon|(x-x_0)}e^{-\epsilon^2t/2} t^{-3/2-\mu}.
\label{dis3}
\end{eqnarray}
The lifetime of a bubble of initial size $x_0$ created at time
$t=0$ follows from Eq. (\ref{dis3}) by calculating the first
passing time density $W(t)$ in Eq. (\ref{def}). Using the
Fokker-Planck equation (\ref{fp2}) we also have more conveniently
\begin{eqnarray}
W(t)=\frac{1}{2}\left[\frac{\partial P}{\partial
x}+\left(\frac{2\mu}{x}-2\epsilon\right)P\right]_{x=0},
\label{abs22}
\end{eqnarray}
and we obtain at long times
\begin{eqnarray}
W(t)\propto
(1+2\mu)x_0^{1+2\mu}e^{|\epsilon|x_0}e^{-\epsilon^2t/2}t^{-3/2-\mu}.
\label{abs31}
\end{eqnarray}

\begin{figure}
\includegraphics[width=8.6cm]{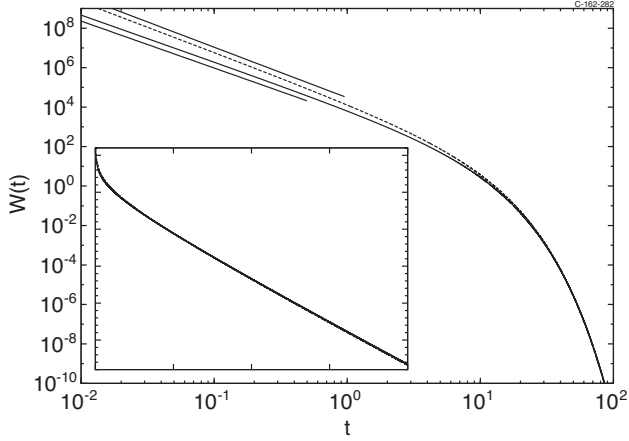}
\caption{Bubble lifetime distribution $W(t)$ from
Eq.~(\ref{abs31}), with $\epsilon=-1/2$, $x_0=5$, and $c=1.76$
(full line) and $2.12$ (dashed). The initial power-law behavior
with slopes -2.38 and -2.56 is indicated by the straight lines.
Inset: $\log$ versus linear scale, emphasizing the exponential
decay for long times.} \label{fig6}
\end{figure}

In Fig.~\ref{fig6} we have depicted the bubble lifetime
distribution $W(t)$ below $T_m$ for $\epsilon = -1/2$.

%
\subsubsection{At the transition $T=T_m$ ($\epsilon=0$)}
\begin{figure}
\includegraphics[width=8.6cm]{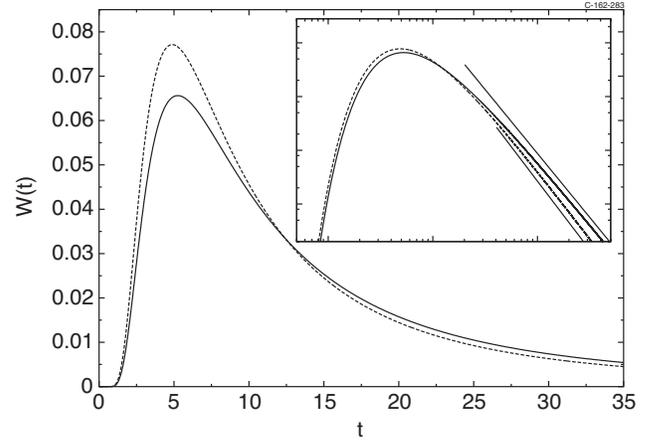}
\caption{Bubble lifetime distribution $W(t)$ from Eq.~(\ref{abs4})
for $T=T_m$, $x_0=5$, as well as $c=1.76$ (full line) and $c=2.12$
(dashed line). Inset: $\log$-$\log$ plot of the power-law behavior
at long $t$, with slopes $-2.38$ and $-2.56$, as indicated by the
straight lines.} \label{fig7}
\end{figure}

At the transition temperature $T=T_m$ for $\epsilon=0$ the Coulomb
term is absent and we have a free particle subject to the
centrifugal barrier $\mu(\mu+1)/2x^2$. In this case the
eigenfunctions are given by the Bessel function \cite{Lebedev72}
\begin{subequations}
\begin{eqnarray}
&&\Psi_k(x)=(kx)^{1/2}J_{1/2+\mu}(kx), \label{bessel}
\\
&&E_k=\frac{k^2}{2}, \label{ek}
\end{eqnarray}
\end{subequations}
where orthogonality and completeness follow from the
Fourier-Bessel integral \cite{Lebedev72}
\begin{eqnarray}
f(x)=\int_0^\infty kJ_\nu(kx)dk\int_0^\infty yJ_\nu(ky)f(y)dy
\label{fb}
\end{eqnarray}
By insertion into Eq.~(\ref{dis}) we obtain the distribution
\begin{eqnarray}
\nonumber
P(x,x_0,t)&=&\frac{x_0^{1/2+\mu}}{x^{\mu-1/2}}\\
&&\times\int_0^\infty dk e^{-k^2t/2}kJ_{1/2+\mu}(kx)J_{1/2+\mu}(kx_0),
\nonumber \label{dis4}
\end{eqnarray}
or, by means of the identity \cite{Lebedev72}
\begin{equation}
\int_0^\infty e^{-tx^2}J_p(ax)J_p(bx)xdx=
\frac{1}{2t}e^{-(a^2+b^2)/4t}I_p\left(\frac{ab}{2t}\right),
\end{equation}
the explicit expression
\begin{eqnarray}
P(x,x_0,t)&=&\left(\frac{x_0}{x}\right)^{\mu}(xx_0)^{1/2} t^{-1}
e^{-(x^2+x_0^2)/2t} \nonumber
\\
&&\times I_{1/2+\mu}(xx_0/t). \label{dis5}
\end{eqnarray}
Here, $I_\nu(z)$ is the Bessel function of imaginary argument
\cite{Lebedev72}. From Eq. (\ref{dis5}) we also infer, using Eq.
(\ref{abs22}) the first passage time distribution
\begin{eqnarray}
W(t)=\frac{2x_0^{1+2\mu}}{\Gamma(1/2+\mu)}e^{-x_0^2/2t}(2t)^{-3/2-\mu},
\label{abs4}
\end{eqnarray}
in accordance with Eq. (\ref{abs31}) for $\epsilon=0$. In
Fig.~\ref{fig7} we show the first passage time distribution
(\ref{abs4}) for two different critical exponents $c$. Note that
the power-law exponent $-3/2-\mu=-3/2-c/2$ is identical to the
result reported in Ref.~\cite{Bar07}.
%
\subsubsection{Long times for $T>T_m$}
Above the transition temperature for $\epsilon>0$ the Coulomb
potential $-\mu\epsilon/x$ is attractive and can trap a series of
bound states. In the long time limit the lowest bound state
controls the behavior of $P$. According to Eqs. (\ref{ham4}) and
(\ref{eigen}) the lowest bound state $\Psi_1$ with eigenvalue
$E_1<\epsilon^2/2$ must satisfy the eigenvalue equation
\begin{eqnarray}
\left[-\frac{1}{2}\frac{d^2\Psi_1}{dx^2}
+\frac{\mu(\mu+1)}{2x^2}-\frac{\mu\epsilon}{x}+
\frac{\epsilon^2}{2}\right]\Psi_1=E_1\Psi_1. \label{eigen22}
\end{eqnarray}
For $x\rightarrow\infty$ we have
$-(1/2)\Psi_1^{\prime\prime}=(E_1-\epsilon^2/2)\Psi_1$ and
$\Psi_1$ must fall off exponentially, $\Psi_1\sim\exp(-\lambda
x)$, $\lambda=(2E_1-\epsilon^2)^{1/2}$. For $x\rightarrow 0$ we
have $-(1/2)\Psi_1^{\prime\prime}+(\mu(\mu+1)/2x^2)\Psi_1\sim 0$
and we infer $\Psi_1\sim x^{1+\mu}$. Consequently, searching for a
nodeless bound state of the form $\Psi_1\sim
x^{1+\mu}\exp(-\lambda x)$ we readily obtain the normalized lowest
level
\begin{subequations}
\begin{eqnarray}
&&\Psi_1(x)=Ax^{1+\mu}e^{-\mu\epsilon x/(1+\mu)}, \label{bound}
\\
&&A^2=\frac{(2\mu\epsilon/(\mu+1))^{2\mu+3}}{\Gamma(2\mu+3)},
\label{norm}
\end{eqnarray}
\end{subequations}
with corresponding eigenvalue
\begin{eqnarray}
E_1=\frac{\epsilon^2}{2}
\left(1-\left(\mu/(\mu+1)\right)^2\right). \label{e1}
\end{eqnarray}
The maximum of the bound state is located at
$(\mu+1)^2/\mu\epsilon\sim 1/(T-T_m)$ and thus recedes to infinity
as we approach the melting temperature. From Eq. (\ref{dis}) we
thus obtain after some reduction
\begin{eqnarray}
P(x,x_0,t)&=&A^2 x
x_0^{1+2\mu}e^{(\epsilon/(1+\mu))(x-x_0(1+2\mu))} \nonumber
\\
&&\times e^{-\epsilon^2(1+2\mu)t/2(1+\mu)^2}. \label{dis6}
\end{eqnarray}
Above $T_m$ the bubble size, on average, increases in time until full
denaturation is reached. In terms of the free energy plot in Fig.~\ref{fig1}b
this corresponds to a Kramers escape across the (soft) potential barrier
(corresponding to a nucleation process).
This implies that the transition probability $P(x,x_0,t)$ from an
initial bubble size $x_0$ to a final bubble size $x$ must vanish in
the limit of large $t$. According to Eq. (\ref{dis6}) $P(x,x_0,t)$
decays exponentially,
\begin{eqnarray}
P(x,x_0,t)\propto e^{-t/\tau} \label{dis7}
\end{eqnarray}
with a time constant given by
\begin{eqnarray}
\tau=\frac{2(1+\mu)^2}{(1+2\mu)\epsilon^2}\propto|T-T_m|^{-2}
\label{time}
\end{eqnarray}
%
\subsection{Exact results}
The eigenvalue problem given by Eqs. (\ref{ham4}) and (\ref{eigen})
\begin{eqnarray}
\left[-\frac{1}{2}\frac{d^2\Psi}{dx^2}
+\frac{\mu(\mu+1)}{2x^2}-\frac{\mu\epsilon}{x}+
\frac{\epsilon^2}{2}\right]\Psi=E\Psi, \label{eval}
\end{eqnarray}
has the same form as the differential equation satisfied by the
Whittaker function $w$ \cite{Gradshteyn65},
\begin{eqnarray}
-\frac{d^2w}{dx^2}+
\left(\frac{1}{4}-\frac{\lambda}{z}-\frac{1/4-m^2}{z^2}\right)w=0,
\end{eqnarray}
with the identifications $z=2\kappa x$,
$\lambda=\mu\epsilon/\kappa$, $m=1/2+\mu$, and
$E=\epsilon^2/2-\kappa^2/2$. Incorporating the absorbing state
condition $\Psi(0)=0$ and using an integral representation for the
Whittaker function $w$ \cite{Gradshteyn65} we obtain the solution
\begin{eqnarray}
\nonumber
\Psi(x)&\propto& (2\kappa x)^{1+\mu} e^{-\kappa x}\\
&&\times\int_0^\infty e^{-2\kappa xt}
t^{\mu(1-\epsilon/\kappa)}(1+t)^{\mu(1+\epsilon/\kappa)}dt.
\nonumber
\\
\label{int}
\end{eqnarray}
In the bound state case for $\epsilon>0$ the parameter $\kappa>0$
and the bound state spectrum is obtained by terminating the power
series expansion of Eq. (\ref{int})\cite{Gradshteyn65},
\begin{eqnarray}
\nonumber
\Psi(x)&\propto& (2\kappa x)^{1+\mu} e^{-\kappa x}\\
&&\times\Phi(1+\mu(1-\epsilon/\kappa),2(1+\mu);2\kappa x), \label{pow1}
\end{eqnarray}
with the polynomial
\begin{eqnarray}
\nonumber
\Phi(\alpha,\gamma;z)&=&1+\frac{\alpha}{\gamma}\frac{z}{1!}+
\frac{\alpha(\alpha+1)}{\gamma(\gamma+1)}\frac{z^2}{2!}\\
&&+\frac{\alpha(\alpha+1)(\alpha+2)}{\gamma(\gamma+1)(\gamma+2)}
\frac{z^3}{3!}.\label{pow2}
\end{eqnarray}
Simple algebra then yields the spectrum
\begin{eqnarray}
\kappa=\epsilon\frac{\mu}{\mu+n},~~n=1,2,\cdots \label{eval2}
\end{eqnarray}
and associated eigenfunctions
\begin{eqnarray}
\Psi\propto x^{1+\mu}e^{-\kappa x}\times\text{polynomial},
\label{eigen2}
\end{eqnarray}
the lowest state and eigenfunctions given by Eqs. (\ref{bound})
and (\ref{e1}).
\section{\label{discussion}Discussion}
In typical experiments measuring fluorescence correlations of a
tagged base pair bubble breathing can be measured on the level of
a single DNA molecule \cite{Bonnet98,Altan-Bonnet03}. The correlation
function $C(t)$ is proportional to the integrated survival
probability, i.e.,
\begin{eqnarray}
C(t)\propto\int_0^LP(x,x_0,t)dx, \label{corr}
\end{eqnarray}
where $L$ is the chain length. From the definition of the first
passage time distribution in Eq. (\ref{def}) we also have
\begin{eqnarray}
C(t)=1-\int_0^tW(t')dt'. \label{corr2}
\end{eqnarray}
%
\subsection{Below $T_m$ for $\epsilon<0$}
Below the melting temperature $T_m<0$ we obtain from Eq.
(\ref{abs31})
\begin{eqnarray}
C(t)= 1-x_0^{1+2\mu}e^{|\epsilon|x_0}
\int_0^te^{-\epsilon^2t'/2}(t')^{-3/2-\mu} dt', \label{corr3}
\end{eqnarray}
or in terms of the incomplete Gamma function
$\gamma(\alpha,x)=\int_0^x e^{-t}t^{\alpha-1}dt$ \cite{Lebedev72}
\begin{eqnarray}
\nonumber
C(t)&=&1-x_0^{1+2\mu}e^{|\epsilon|x_0}(\epsilon^2/2)^{1/2+\mu}\\
&&\times\gamma(-1/2-\mu,\epsilon^2t/2).
 \label{corr4}
\end{eqnarray}
Using $\gamma(\alpha,x)=\Gamma(\alpha)-x^{\alpha-1}e^{-x}$ for
$x\rightarrow \infty$ we have for large $t$
\begin{eqnarray} C(t)=\text{const.}+
x_0^{1+2\mu}\epsilon^{-2}e^{|\epsilon|x_0}t^{-3/2-\mu}
e^{-\epsilon^2 t/2}. \label{corr5}
\end{eqnarray}
We note that the basic time scale of the correlations is set by
$\epsilon^{-2}\propto (T_m-T)^{-2}$. As we approach $T_m$ the time
scale diverges like $(T_m-T)^{-2}$.

For $t\ll\epsilon^{-2}$ the correlations show a power law behavior
\begin{eqnarray} C(t)=\text{const.}+
C(t)\propto t^{-3/2-\mu}\text{(mod a const.)}, \label{corr6}
\end{eqnarray}
with scaling exponent $-3/2-\mu=-3/2-c/2$. Here $3/2$ originates from unbiased
bubble size random walk whereas the contribution $\mu=c/2$ is
associated with the entropy loss of a closed polymer loop.

At long times $t\gg\epsilon^{-2}$ the correlations fall off
exponentially
\begin{eqnarray} C(t)=\text{const.}+
C(t)\propto e^{-\epsilon^2t/2}\text{(mod a const.)}. \label{corr7}
\end{eqnarray}
The size of the time window showing power law behavior increases
as $T_m$ is approached. This corresponds to the critical slowing down on
denaturation, as already observed in Ref.~\cite{Ambjorn06} numerically, and in
Ref.~\cite{Bicout04} in absence of the critical exponent $c$ due to polymeric
interactions.

\begin{figure}
\includegraphics[width=1.\hsize]{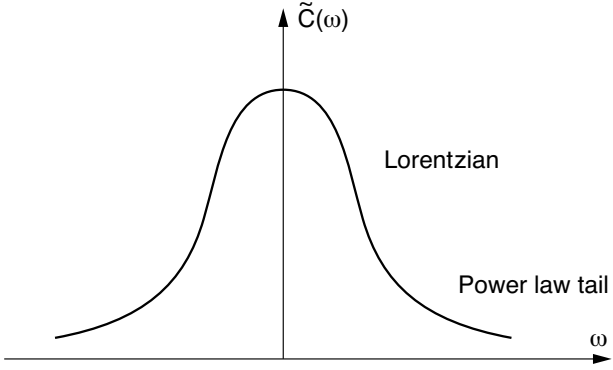}
\caption{The structure function $\tilde C(\omega)$. For
$|\omega|\ll\epsilon^2$ the structure function has a Lorentzian
line shape; for $|\omega|\gg\epsilon^2$ it exhibits power law
tails.} \label{fig8}
\end{figure}

In frequency space the structure function  is given by
\begin{eqnarray}
\tilde C(\omega)=\int e^{i\omega t}C(t)dt. \label{struc}
\end{eqnarray}
By means of a simple scaling argument we infer that $\tilde
C(\omega)$ has a Lorentzian line shape for $|\omega|\ll\epsilon^2$
crossing over to power law tails for $|\omega|\gg\epsilon^2$.
\begin{subequations}
\begin{eqnarray}
&&\tilde C(\omega)\sim
x_0^{1+2\mu}e^{|\epsilon|x_0}\frac{1}{\omega^2+
(\epsilon^2/2)^2}~~\text{for}~~|\omega|\ll\epsilon^2~~~~~~
\label{struc2}
\\
&&\tilde C(\omega)\sim
x_0^{1+2\mu}e^{|\epsilon|x_0}\frac{1}{\epsilon^2}
|\omega|^{1/2+\mu}~~~~~\text{for}~~|\omega|\gg\epsilon^2
\label{struc3}
\end{eqnarray}
\end{subequations}
In Fig.~\ref{fig8} we have depicted the structure function $\tilde
C(\omega)$.
%
\subsection{At $T_m$ for $\epsilon=0$}
At the transition temperature $T_m$ the exact expression for the
first passage time distribution is given by Eq. (\ref{abs4}).
Using Eq. (\ref{corr2}) for $C(t)$ we then obtain
\begin{eqnarray}
C(t)=1-\frac{\Gamma(1/2+\mu,x_0^2/2t)}{\Gamma(1/2+\mu)},
\label{corr8}
\end{eqnarray}
where $\Gamma(\alpha,x)=\int_x^\infty e^{-t}t^{\alpha-1}dt$ is the
incomplete Gamma function \cite{Lebedev72}.

At short times we have
\begin{eqnarray}
C(t)=1-\frac{(x_0^2/2)^{\mu-1/2}}{\Gamma(1/2+\mu)}
t^{1/2-\mu}e^{-x_0^2/2t}, \label{corr9}
\end{eqnarray}
whereas for $t\rightarrow\infty$
\begin{eqnarray}
C(t)=
\frac{2(x_0^2)^{1/2+\mu}}{(1+2\mu)\Gamma(1/2+\mu)}t^{-1/2-\mu}.
\label{corr10}
\end{eqnarray}
The correlation function thus exhibits a power law behavior with scaling
exponent $-1/2-\mu=-1/2-c/2$, as obtained from a different argument in
Ref.~\cite{Bar07}. Correspondingly, the structure function
$\tilde C(\omega)$ has the form
\begin{eqnarray}
\tilde C(\omega)\propto x_0^{1+2\mu}|\omega|^{\mu-1/2}.
\label{struc4}
\end{eqnarray}
%
\subsection{Above $T_m$ for $\epsilon>0$}
\begin{figure}
\includegraphics[width=1.\hsize]{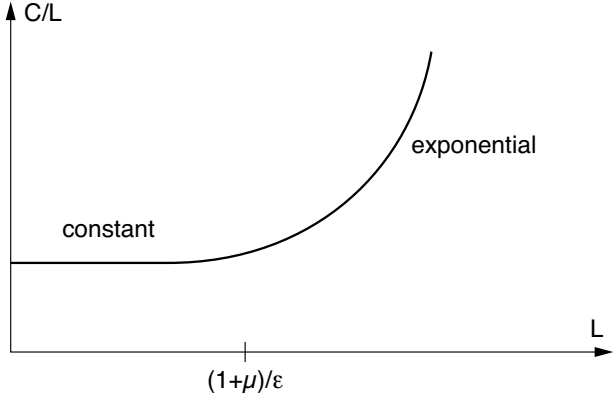}
\caption{We depict $C/L$ as a function of $L$. For
$L\ll(1+\mu)/\epsilon$ the correlations depends linearly on $L$;
for $L\gg(1+\mu)/\epsilon$ the correlations increase exponentially
as a function of $L$.} \label{fig9}
\end{figure}

Above $T_m$ ($\epsilon>0$) the DNA chain eventually fully
denatures and the correlations diverge in the thermodynamic limit.
We can, however, at long times estimate the size dependence for a
chain of length $L$. From the general expression (\ref{dis}) we
find
\begin{eqnarray}
C(t)\simeq e^{-\epsilon x_0} x_0^\mu\sum_n e^{-E_nt}\Psi_n(x_0)
\int_0^L e^{\epsilon x}x^{-\mu}\Psi_n(x)dx. \nonumber\\
~ \label{corrgen}
\end{eqnarray}
At long times the lowest bound state dominates the expression.
Inserting $\Psi_1$ and $E_1$ from Eqs. (\ref{bound}), (\ref{norm}),
and (\ref{e1}) and performing the integration over $x$ we obtain
\begin{eqnarray}
\nonumber &&C(t)\propto A^2e^{-\epsilon x_0(2\mu+1)/(\mu+1)}
e^{-\epsilon^2((\mu+1/2)/(\mu+1)^2)t}x_0^{1+2\mu}\\
&&\times (1+\mu)\epsilon^{-2}
\left[1+(L\epsilon/(1+\mu)-1)e^{\epsilon L/(1+\mu)}]\right].
\label{corr11}
\end{eqnarray}
The correlations decay exponentially with time constant $\sim
\epsilon^{-2}(\mu+1)^2/(2\mu+1)$. In frequency space the structure
function has a Lorentzian lineshape of width
$\sim\epsilon^2(2\mu+1)/(\mu+1)^2$, and for the size dependence one
obtains
\begin{equation}
C(t)\sim\left\{\begin{array}{ll}
L e^{\epsilon L/(1+\mu)}, & \mbox{for } \epsilon L/(1+\mu)\gg 1,\\[0.2cm]
L\epsilon/(1+\mu), & \mbox{for } \epsilon L/(1+\mu)\ll 1
\end{array}\right..
\label{corr12}
\end{equation}
Note that close to $T_m$ the correlation function $C(t)\propto L$.
In Fig.~\ref{fig9} we depict in a plot of $C/L$ vs. $L$ the size
dependence of the correlation function.
\begin{figure}
\includegraphics[width=8.6cm]{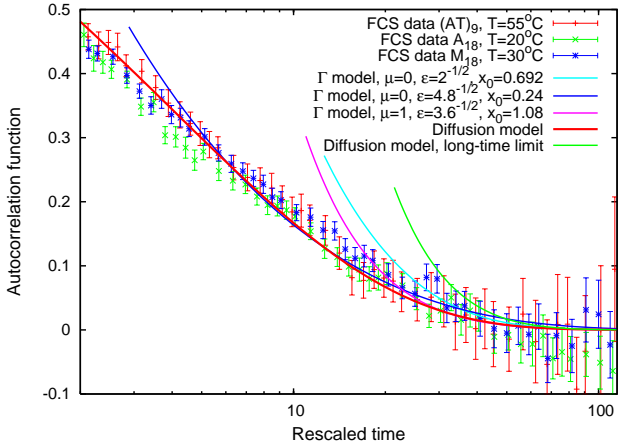}
\caption{(Color online). Drift-diffusion model and experimental
data from Ref.\cite{Altan-Bonnet03} compared to the $\Gamma$
model, for various parameters. The curve for $\mu=0$ and
$\epsilon=1/\sqrt{2}$ exactly matches the long-time behavior from
Ref.~\cite{Altan-Bonnet03}.} \label{fig10}
\end{figure}

\subsection{Comparison to experimental data}

Below the melting temperature $T_m$, DNA breathing can be monitored on
the single DNA level by fluorescence correlation spectroscopy
\cite{Ambjorn06,Ambjorn07,Altan-Bonnet03}. In the FCS experiment from
Ref.~\cite{Altan-Bonnet03}, a DNA construct of the form
\begin{equation}
\begin{array}{l}
\mbox{\texttt{\textcolor{white}{5'GGCGCCCATATATATATA}F\textcolor{white}{ATATATATGCGCTT}}}\\[-0.05cm]
\mbox{\texttt{5'\textcolor{white}{GGCGCCCATATATATATA}|\textcolor{white}{ATATATATGCGC}T\textcolor{white}{T}}}\\[-0.05cm]
\mbox{\texttt{\textcolor{white}{5'}GGCGCCCATATATATATATATATATATGCGC\textcolor{white}{T}T}}\\[-0.05cm]
\mbox{\texttt{\textcolor{white}{3'}CCGCGGGTATATATATATATATATATACGCG\textcolor{white}{T}T}}\\[-0.05cm]
\mbox{\texttt{3'\textcolor{white}{GGCGCCCATATATATAT}|\textcolor{white}{TATATATATGCGC}T\textcolor{white}{T}}}\\[-0.05cm]
\mbox{\texttt{\textcolor{white}{5'GGCGCCCATATATATAT}Q\textcolor{white}{TATATATATGCGCTT}}}
\end{array}
\end{equation}
was employed. Here, a bubble domain consisting of weaker AT base
pairs are clamped by stronger GC base pairs. On the right, a short
loop consisting of four T nucleotides is introduced. The
fluorophore (F) and quencher (Q) are attached to T nucleotides as
shown. With the highest probability, a bubble will form in the
AT-bubble domain. As the bubbles consist of flexible
single-strand, in an open bubble the fluorophore and quencher move
away from each other, and fluorescence occurs. Once in the focal
volume of the FCS setup, bubble opening and closing corresponds to
blinking events in the signal, whose correlation function
(corrected for the diffusion in and out of the focal volume) are
shown in Fig.~\ref{fig10}. Three different bubble domains with
changing sequence were used to check that potential secondary
structure formation does not influence the breathing dynamics,
confirming the picture of base pair-after-base pair zipping and
unzipping. The figure shows examples from all three constructs,
underlining the data collapse already observed in
Ref.~\cite{Altan-Bonnet03}.

The theoretical lines shown in Fig.~\ref{fig10} correspond to the
biased diffusion model introduced in the original article
\cite{Altan-Bonnet03}. While the full solution of this diffusion
model fits the data well over the entire window, the long time
expansion demonstrates the rather weak convergence of the
expansion. In Fig.~\ref{fig10} we also included our asymptotic
solution (\ref{corr4}) for the autocorrelation function, for
various parameters. Good agreement with the data is observed.

\section{\label{summary}Summary and Conclusion}
In this paper we have analyzed the breathing dynamics of thermally
induced denaturation bubbles forming spontaneously in
double-stranded DNA. We have shown that the Fokker-Planck equation
can be analyzed from two points of view: i) In the weak noise or
low temperature limit a canonical phase space approach interprets
the stochastic dynamics in terms of a deterministic 'classical'
picture and gives by simple estimates access to the long time
dynamics. In particular, we deduce that the dynamics at the
transition temperature is characterized by power law behavior with
scaling exponent depending on the entropic term. ii) In the
general case we show that the Fokker-Planck equation can be mapped
onto the imaginary time Schr{\"o}dinger equation for a particle in
a Coulomb potential. The low temperature region below the
transition temperature then corresponds to the continuum states of
a repulsive Coulomb potential, whereas the region above $T_m$ is
controlled by the lowest bound state in an attractive Coulomb
potential. The mapping, moreover, allows us to calculate the
distribution of bubble lifetimes and the associated correlation
functions, below, at, and above the melting temperature of the DNA
helix-coil transition. Finally, at the melting transition, the DNA
bubble-breathing was revealed to correspond to a one-dimensional
finite time singularity.

The analysis reveals non-trivial scaling of the first passage time
density quantifying the survival of a bubble after its original
nucleation. The associated critical exponent depends on the
parameter $\mu=c/2$ stemming from the entropy loss factor of the
flexible bubble. The first passage time distribution
and correlations depend on the difference $T/T_m-1$, and
therefore explicitly on the melting temperature $T_{\mathrm{m}}$
(and thus the relative content of AT or GC base pairs). We also
obtained the critical dependence of the characteristic time scales
of bubble survival and correlations on the difference $T-T_m$. The
finite size-dependence of the correlation function was recovered,
as well.

The mapping of the of DNA-breathing onto the quantum Coulomb problem
provides a new way to investigate its physical properties, in
particular, in the range above the melting transition, $T>T_m$. The
detailed study of the DNA bubble breathing problem is of particular
interest as the bubble dynamics provides a test case for new
approaches in small scale statistical mechanical systems where the
fluctuations of DNA bubbles are accessible on the single molecule
level in real time.

\acknowledgments

Discussions with T. Ambj{\"o}rnsson, S. K. Banik, O. Krichevsky,
and A. Svane are gratefully acknowledged. We thank O. Krichevsky
for providing the fluorescence correlation data used in
Fig.~\ref{fig10}. The present work has been supported by the
Danish Natural Science Research Council, the Natural Sciences and
Engineering Research Council (NSERC) of Canada, and the Canada
Research Chairs program.

\end{document}